\documentstyle[12pt,aasms4]{article}
\lefthead{Beauchamp, D. \& Hardy, E.}
\righthead{Metallicity Indices In Galaxies...} 
\begin{document}

\title{The Stellar Populations of Spiral Disks. I. A New Observational
Approach: Description of the Technique and Spectral
Gradients for the Inter--Arm Regions of NGC~4321 (M~100)}

\author{Dominique Beauchamp and Eduardo Hardy}

\affil{D\'epartement de physique, Universit\'e Laval,\\ 
Observatoire du mont M\'egantic\\
{\it beaucham@phy.ulaval.ca, hardy@phy.ulaval.ca}}

\authoraddr{beaucham@phy.ulaval.ca, hardy@phy.ulaval.ca}


\begin{abstract}
We describe an imaging method that makes use of interference filters to
provide integrated stellar spectral indices for spiral disks to faint
surface brightness limits. We use filters with bandpasses ${\sim 60
{\rm \AA\ }}$FWHM, centered on the Mg and Fe features ($\lambda
\lambda $5176 \AA \ and 5270 \AA\ respectively) allowing the
determination of the spatial distribution of the Lick indices ${\rm
Mg_{2}}$ and ${\rm Fe5270}$. These two indices have been extensively
modeled by different groups and used in the past mainly for the study
of elliptical galaxies, bulges, and globular clusters. Azimuthal
integration of the underlying smooth stellar signal, after removal of
the signature of the spiral arms and associated extreme pop. I
structures, provides measurements of these spectral indices useful to
radial distances where the surface brightness of the galaxy reaches
$\sim$ 24$\mu_V$ .

As a first example of this technique and its possibilities we conduct a
preliminary study of the SABbc galaxy NGC~4321 (M~100). We present
spectral gradients for the inter--arm stellar population to about 4
exponential scale lengths. There is some evidence for a discontinuity
in the run of the ${\rm Mg_{2}}$ index near corotation, which we
interpret as evidence for bar-- driven secular evolution. There is also
evidence that Mg is overabundant with respect to Fe in the inner
regions of the projected image.

\end{abstract}

\section{Introduction}

Spectral gradients contain a wealth of information on the stellar
populations of galaxies, and therefore on their formation and
evolution. Studies of gradients in ellipticals and S0 galaxies, where
the surface brightness remain high to large radii, are by now common
(Couture \& Hardy 1988\markcite{CH88}, Gorgas et al. 
1990\markcite{GEA90}, Davidge 1992\markcite{D92}, Carollo et al. 
1993\markcite{CDB93}, Davies et al. 1993\markcite{DSP93}
Casuso et al. 1996\markcite{CVPB96}). For spiral galaxies, especially
late-type ones, chemical gradients have been under study for a long
time using the information provided by the bright emission lines of
their \ion{H}{2} regions, for which the physics is well
understood. \ion{H}{2} regions have indeed provided $\alpha$-element
abundances for the gas-phase of spirals to large radial distances
(Belley \& Roy 1992\markcite{BR92}, Scowen et al. 
1992\markcite{SDH92}, Martin \& Roy 1994\markcite{MR94}, Zaritsky 
et al. 1994\markcite{ZKH94}). A number of recent investigations
on the integrated spectra of spiral {\em bulges} provide information on
their age and metallicity structure, much as has been done with
ellipticals (Jablonka et al. 1996\markcite{JMA96}; Idiart et
al. 1996\markcite{IFPDC96}, Davidge 1997\markcite{D97}). Little is
known however about the spectral properties of the {\em stellar} disks of
external spirals (but see Monteverde et al., 1997 for a study of
abundances of O stars in M33, and Brewer, Richer \& Crabtree, 1995 for 
an indirect estimate of the stellar metallicity gradient in M31 from C to M 
stars ratios).

The integrated light of the inter--arm regions of spiral disks
contains a light-weighted chemical signature complementary to that of
\ion{H}{2} regions. We observe in stellar disks a time average over
their chemical history, as opposed to the accummulated effect on the
extreme population I component revealed by the properties of the
excited gas. In addition, the chemistry of stellar disks may contain
valuable information on the dynamical instabilities and secular
evolution introduced by common non-axisymmetric perturbations such as
oval structures and bars (Friedli \& Benz 1995, Courteau et
al. 1996\markcite{CDJB96}). Furthermore, and this is an issue we will
explore briefly in this paper, it is possible to separate in the integrated
light of stellar disks the contribution of the $\alpha$-elements from
that of the Fe-group elements, something that has not yet been possible for
\ion{H}{2} regions. Indeed, stellar iron features, such as Fe5270 and
Fe5335, are commonly used as indicators for species formed in the
low-mass stellar progenitors of SNeI, although some Fe is also formed in
SNeII. Magnesium features such as the one centered at $\lambda $ 5176
\AA, are indicators of $\alpha $-elements formed in the SNeII
explosions in which high-mass stars end their life (see Worthey et al.
1992\markcite{WFG92}). One must keep in mind however that all
integrated stellar spectral indices used to derive abundances contain also
an age dependence that must be taken into account.

Observing the spectral signature of a stellar disk with reasonable S/N
ratios is a difficult task as its surface brightness, which decreases
exponentially with radial distance from the center, reaches very low
levels quickly. Indeed, for most late-type spirals, with exponential
length scales smaller than 5 kpc, the surface brightness at 10 kpc is fainter
than 24$\mu_{B}$, well below that of the night sky. The long-slit
spectroscopic technique employed in the study of ellipticals becomes
unsuitable for spirals even for 4m-class telescopes, especially at low
light levels where accurate sky-subtraction is critical. The use of
modern large-format imaging CCD detectors combined with large fields
of view and well-tested imaging analysis techniques provides a
useful alternative, as shown in this investigation.

In this paper, the first of a series, we describe an imaging method
that uses relatively narrow ($\sim$ 60 \AA ) filters designed to
isolate the Mg and Fe spectral features indicated above. This
technique which allows azimuthal integration of the signal over the
surface of the disk has not been used previously on spirals, and we
will show that it is capable of providing good S/N ratios to large
radial distances. We concentrate here heavily on technical issues
leaving most aspects of the calibration of the spectral indices in
terms of actual abundances for later publications. Results for the
SABbc galaxy NGC~4321 (M~100), a galaxy which has been extensively
studied, and which we used as our test--bench, will however be presented.

The following simple order--of--magnitude computation shows the 
dramatic increase in S/N afforded by the present method when compared to 
the conventional method of long--slit spectroscopy, especially at large
radial distances. We are here setting aside the obvious multiplexing
advantage of a spectrograph.  Consider the case of M~100, and a
section at $R_{25}$ = 222\arcsec . Let us assume that we are using a
grating that provides (as in \S 3) a resolution of 34 \AA\ FWHM,
compatible with the indices studied here, and corresponding to a slit width
of 6\farcs 1 . Let us further assume that we integrate the spectrum
along the slit within a radial interval of 16\arcsec\ providing a
resolution element of $\sim $ 100 $\sq \arcsec\ $. The surface of an
annulus of the same radial width and inner radius is $\sim $ 2.3
$\times$ 10$^4$ $\sq \arcsec $. In other words, if we concentrate on
{\em one} spectral feature, azimuthal integration of the annulus using
images taken through a narrow band-pass filter provides the equivalent
of a signal about 230 times larger than that obtained from a slit
observation, for the same spectral and radial resolution.  For a low
readout noise CCD, this translates into an increase in the effective
S/N of the measurement {\em per filter} approaching a factor 15. The
total readout noise will however increase as a result of the increased
number of summed pixels, but the precision of the sky sampling will be
higher, as will be the effective quantum efficiency of a CCD + filter
combination which is at least 2--3 times that of a grating spectrograph. 
Furthermore, as discussed in  \S 2,  three filters are needed per spectral 
index (one for the feature and two for the continua), thus increasing the 
total exposure time in the imaging method. When all of this is taken into
consideration the expected improvement in S/N for equal pixel binning
and exposure time is still close to a very substantial factor 10. The relative 
improvement is, of course, larger at large radial distances where the S/N 
ratio per unit surface is lowest. For edge--on spirals our argument is 
certainly less compelling, as they would exhibit a higher projected surface 
brightness along the slit. Yet, increased internal absorption and the 
inhability to isolate the underlying smooth stellar disk from the arm 
component are important limitations associated with the study of edge--on 
spirals.

The use of an interference filter--CCD combination to isolate spectral
features in the integrated stellar light of galaxies is not new. For
example, Thomsen \& Baum (1989\markcite{TB89} and references 
therein), have used it to examine the Mg behavior within ellipticals in the 
Coma cluster of galaxies. We use such a technique here on spiral disks
for the first time and pool together the imaging data obtained through
different bandpasses, including wideband ones, to separate the
information belonging to the inter--arm regions of nearly face--on
galaxies from that of the spiral arms.

The interference filter set described below was designed for a mean
target velocity close to that of the Virgo cluster. At that distance we have
excellent spatial resolution per flux unit, atmospheric seeing does not
become an important problem, and our bandpasses do not become
contaminated by any strong sky emission line. We have chosen here
nearly face-on galaxies for the reasons already discused and to increase the 
number of available pixels (see \S 4.1.3), and minimize rotational effects, 
but it will certainly be very rewarding in the future to observe edge-on 
galaxies with the same technique in order to explore the disk-halo 
connection. Our use of a focal reducer with a large field of view rendered 
the determination of the sky level of our images feasible even at the 
relatively short distances of our sample.

In subsequent publications we will combine this observational
technique, as applied to a larger sample of galaxies, with the results
of spectral synthesis, of models of chemical evolution of spiral
galaxies, and of numerical simulations. When combined with observations 
of chemical gradients derived from \ion{H}{2} regions, these results will 
be valuable in improving our understanding of spiral galaxies.

This paper is organized as follows. In \S 2 we discuss the method and
in particular the filter system adopted; we describe the observations
in \S 3, the reduction techniques and calibrations in \S 4, and in \S 5
we present the first results for M~100 (NGC~4321).

\section{The Method}

We have chosen to work on two spectral features: the ${\rm Mg_{2}}$
molecular feature, and the ${\rm Fe5270}$ absorption line (Faber et
al. 1985\markcite{FFBG85}, and references therein (hereafter FFBG85),
Burstein et al. 1984\markcite{BFGK84}; Worthey et al. 
1992\markcite{WFG92}). These two indices have been extensively
modeled by different groups (Worthey 1994\markcite{W94}; Chavez et
al. 1995\markcite{CMM95}; Chavez et al. 1996\markcite{CMM96}, Idiart et al 1996\markcite{IFPDC96}) and
should prove useful in describing the radial behavior of chemical
abundances for the two groups of elements mentioned in \S 1. They are
defined within a restricted spectral interval, and their sensitivity
to dust absorption should be small (see \S 4.2.4). Table~1 provides
the spectral information on the 4 filters used, which were built for
us by Andover Co., and which were designed for objects centered at a
velocity of 2000~${\rm km}\:{\rm s}^{-1}$ (at
23~$^{\circ}$C). The top panel of Figure~1 displays the spectrum of a K 
giant, whereas the bottom panel shows the location and transmittance 
curves of the filters. Notice the important caveat that although we have
fully respected the Lick definition for ${\rm Mg_{2}}$ (Burstein et
al. 1984), we have not done so for the Fe5270 feature. In the latter
case we have decided to adopt provisionally a pseudo--${\rm Fe5270}$
index that retains the original Lick Fe absorption bandpass, but uses the 
same continua bandpasses as the ${\rm Mg_{2}}$ feature to define the
pseudo--index ${\rm Fe5270'}$. We did so in order to reduce the total 
number of filters required and save observing time while testing to at
least a first approximation the Fe behavior. As it turned out this
inconsistency is not a fundamental limitation because the
pseudo--index can be well calibrated in terms of the true Lick index,
as we will show in \S 4.

%\placetable{tab:filters}

%\placefigure{fig1}

Because the effective wavelengths of interference filters shift
towards the blue with increasing inclination ($\sim 1$ \AA \ per
degree, up to 5\arcdeg) and decreasing temperature ($\sim 0.2$ \AA \
per $^\circ$C), they must be fine--tuned at the telescope. Given the
widths of the filters (about 60 ${\rm \AA\:}$ FWHM), the average nightly temperature, and the tilting
capabilities at our disposal, we could observe galaxies spanning the approximate velocity interval 1400~${\rm km}\:{\rm s}^{-1}$ -- 2100~${\rm km}\:{\rm s}^{-1}$. This interval was quite appropriate for Virgo spirals. Notice that the strong ${\rm [OII] \: \lambda 5577 \AA\ } $
atmospheric emission line falls to the red of our filters, an
important consideration when dealing with photometry at small surface
brightness levels (see Baum et al. 1986\markcite{BTM86}).

\section{Observations}

All observations presented here were conducted at the 1.6 m telescope
of the Observatoire du mont M\'egantic (OMM). We used the f/2 focal
reducer PANORAMIX built by Dr. J.--R. Roy for imaging, outfitted with 
a special filterholder that allows for variable filter inclinations. 
Spectroscopy was conducted with the standard OMM Boller \& Chivens long--slit
spectrograph. The same 1024 $\times$ 1024 Thomson CCD detector with a
pixel size of 19 \micron \ was employed for both types of
observations. The setups are summarized on Tables~2 and 3. The
relevant properties of M~100, including its distance and scale, are summarized on Table~4 following the usual notation. 

Images obtained through
interference filters provide uncalibrated indices, which must then be
brought into the Lick standard system. This correction is obtained in
two steps via spectroscopic observations. First, spectroscopy of a
number of bright Lick standard stars and some galaxies having
published indices were used to set the calibration slopes that
transform our filter system into the Lick system via numerical
integration of the observed energy distribution within each filter bandpass. 
Since our filter definition for the ${\rm Mg_{2}}$ feature closely matches 
the standard
definition we expected from the onset to obtain a slope close to unit, as was
indeed the case, but this would certainly not be true of the Fe5270
index. The second step involves the computation of the zero point of
the system for each galaxy. Since the indices are formed by combining 
sequences of images taken through different filters under variable 
atmospheric conditions, their zero points  must be
established from spectroscopy of the bright nuclear regions of the target
galaxies so as to observe the entire spectral range required 
simultaneously. These and other aspects of the calibration will be discussed
in detail in the following section.

%\placetable{tab:imgsetup}

%\placetable{tab:specsetup}

%\placetable{tab:4321}

A complete set of biases, dome flats and sky flats was obtained on
each night reaching count levels approximately equal to half the
full--well capacity of the CCD. Dark count rates proved variable, and
a set of small masks near the periphery of the field were used to
measure for every exposure the additive signal composed of the dark
signal plus the diffuse-light component.

\section{Data Reduction}

We reduced all the data, images and long--slit spectra, using
IRAF\footnote{\footnotesize IRAF is distributed by the National
Optical Astronomical Observatory, which is operated by the Association
of Universities for Research in Astronomy, Inc., under contract to the
National Science Foundation.} tasks to subtract the overscan, the
bias, and the dark plus diffuse-light frame, and to divide by the
flatfield. Cosmic rays were rejected by the usual clipping algorithm
applied to multiple images.

Following the correction of all frames for the instrumental signature
we co--added the information after performing the geometrical
registration needed to place all  images on the same coordinate
system. 

\subsection{Obtaining Instrumental Spectral Indices from Images}

All indices were derived from the instrumental integrated fluxes
derived from azimuthal integration within annuli. We retained however
the identity of each pixel within a given annulus in order to be able
to tag it for inclusion or exclusion via a mask function $\aleph
(r,\theta)$ ($=1$ for valid pixels, 0 otherwise). This mask was used
to characterize each pixel as belonging to the disk in the inter--arm
region, or to the spiral arms, and reject foreground and background
contamination; we stress here that the imaging method on near face--on
galaxies allows a direct way of computing $\aleph (r,\theta)$.

The following equations were adopted to compute the indices:
\begin{equation}
{\rm Mg_2}=-2.5\log \left( \frac{{\cal F}_{\rm Mg_2}}{{\cal F}_{\rm
C}} \right) \;\;\;{\rm [mag]}
\end{equation}
\begin{equation}
{\rm Fe5270'}=\Delta\lambda \left( 1- \frac{{\cal F}_{\rm
Fe5270}}{{\cal F}_ {\rm C_{\rm w}}} \right) \;\;\;{\rm [\AA]}
\end{equation}
where the ${\cal F}$'s stand for the total sky-subtracted flux seen
through each filter within a given annulus, as given by the following
expression:
\begin{equation}
{\cal F} = \sum_{i=1}^{n}{\rm pixel}_n = \int_{0}^{2\pi}
\int_{r_0}^{r_0+\Delta r} f(r,\theta)\aleph (r,\theta) dr
d\theta\;\;\;{\rm [e^{-}]}
\end{equation}
In the preceding equations $n$ is the number of pixels inside an
annulus, $r_0$ is the inner radius of this annulus, $\Delta r$ is its
width, $f(r,\theta)$ is the flux at a given pixel. The subscript ${\rm
C}$ indicates the continuum level associated to a given bandpass.
Equation~3 can thus be understood as the sum of the sky-subtracted
counts in all valid pixels inside an annulus for a given filter.  In
the equation for Fe5270$'$, $\Delta\lambda$ is the FWHM of the central
filter (e.g. 55 \AA\ ; see Table 1).

The arithmetic mean of the continua (indicated by the subscript ${\rm
C}$) is used to compute the Mg$_2$ index. But since the Fe5270$'$
feature is much closer to the adopted red continum than to the blue one, a
weighted geometrical mean of the continuum at the feature position 
(indicated by
${\rm C_{\rm w}}$) was adopted for the computation of Fe5270$'$
index. The adopted continuum values are given by:
\begin{equation}
{\cal F}_{\rm C} = \frac{{\cal F}_{\rm BC}+{\cal F}_{\rm
RC}}{2}
\end{equation}
\begin{equation}
{\cal F}_{\rm C_{\rm w}} = {\cal F}^a_{\rm BC} \cdot {\cal F}^b_{\rm
RC}
\end{equation}
where $a=0.184$ and $b=0.816$. These values represent the ratio
of the distance in $\lambda$ between the Fe bandpass and each continuum
bandpasses. Notice that Equation 5 (where $ a + b = 1$ ) is a good 
approximation of Equation 4 for symmetric continuum bandpasses (i.e. 
$a=b$)

The details of the calculations of the errors associated with the
spectral indices are given in Appendix A. These calculations provide a
reasonable approximation to the errors. Some small factors are
nevertheless neglected such as the intrinsic dispersion due to the
spectral gradient within a bin.

\subsubsection{Image Tabulation}

In order to perform the azimuthal integration, the images were
tabulated to produce files containing as entries the flux for each
pixel on each band, and its position (in cartesian coordinates
relative to a corner of the image). Because we wanted to mask
foreground stars and other unwanted objects such as residual cosmic
rays and background galaxies, we produced a mask image which was
included in the main table. It was thus easy to reject all entries
corresponding to these masked pixels.

Once the tabulation was done the coordinates relative to the centroid
of the galaxy were computed:
\begin{equation}
x'=x-x_C\;\;\;{\rm [pixels]}
\end{equation}
\begin{equation}
y'=y-y_C\;\;\;{\rm [pixels]}
\end{equation}
where $(x,y)$ are measured relative to the image, $(x',y')$ to the
centroid, and $(x_C,y_C)$ are the position of the centroid relative to
the image. This corresponds to a simple translation. The coordinate
system is then deprojected to produce a face--on view. Deprojection (which 
involves about 1 million pixels) is done by a simple matrix:
\begin{equation}
\left[ \begin{array}{c} x_\circ \\ y_\circ \end{array} \right] = s
\left[ \begin{array}{cc} \cos P.A./ \cos i & 
\sin P.A. / \cos i \\ -\sin {\rm P.A.} & \cos {\rm P.A.} 
\end{array} \right]
\left[ \begin{array}{c} x' \\ y' \end{array} \right]\;\;\;{\rm ['']}
\end{equation}
where $x_\circ$ and $y_\circ$ are coordinates after deprojection,
$P.A.$ is the position angle, $i$ is the inclination and $s$ is the
image scale in \arcsec /pixel.  For each pixel, the {\em deprojected}
distance to the centroid was then computed:
\begin{equation}
r_\circ=\sqrt{x_\circ^2+y_\circ^2}\;\;\;{\rm ['']}
\end{equation}

The following entries were tabulated after completion of the above
operations: $\cal F_{\rm BC}$, $\cal F_{\rm Mg_2}$, $\cal F_{\rm
Fe5270'}$, $\cal F_{\rm RC}$, $\cal F_{\rm C}$, $\cal F_{\rm
C_w}$ $x$, $y$, $x'$, $y'$, $x_\circ$,
$y_\circ$ and $r_\circ$. To these we added $\cal F_{\rm V'}$ and $\cal F_{\rm I}$ (see \S 4.1.3).

The master table was then segmented into annular bins 10\arcsec \
wide, a good compromise between good radial resolution (requiring
small bins) and good S/N (requiring large ones). For each annulus and
each band the mean, median and standard deviation were then computed
for the {\em mean} radial distance $r$ of the bin, which from then on
will characterize it. This produced a new table containing, for each
bin: $\overline{\cal F_\alpha}$, $\sigma _{\cal F_\alpha}$, ${\rm
med}({\cal F_\alpha})$, where $\alpha$ stands for C, C$_{\rm w}$,
Mg$_2$, Fe5270$'$ and $r_\circ$, and $n$ the number of pixels per bin.
We then computed the ratios of annuli's mean radius to the galaxy's
{\em effective radius ($R_{\rm eff}$)} and {\em isophotal radius
($R_{25}$)} so as to be able to plot all galaxies on the same scale.

\subsubsection{Sky Level Computation}

In order to compute indices the sky level has to be subtracted
accurately from all pixel values as indicated in Equations~1 and 2. The
sky level is an {\em additive} constant and it will obviously
introduce non-linear effects on the computed indices if it is not
correctly removed.  Errors in this quantity will be less important
close to the nucleus where the flux from the object dominates, but
they will grow in importance as the object dims with increasing
radius. At the outer limits where the surface brightness is very low
the sky subtraction is extremely critical, and at $R_{25}$, the ratio
sky/galaxy reaches 100.

We tested a number of methods to compute the mode of the distribution
of sky pixels. We fitted gaussians to the histogram and also computed
the mean and the median on the most distant annuli ($450\arcsec \leq R
\leq 520\arcsec$). The mean-median difference is an estimator of the
skewness of distribution (i.e, Wadsworth, 1990\markcite{W90}, pp. 2.8) 
resulting from a galaxy halo, bad (cold and hot) pixels and cosmic
rays that have been missed by the rejection algorithm (as is often the
case with grazing rays). We concluded that the best way to proceed was
to compute the median of all pixels located further away than a number $x$ of isophotal radii ($R_{25}$, from RC3, de Vaucouleurs et
al. 1991\markcite{RC3}). At this distance from the centroid, the
distribution in intensity is close enough to a gaussian so we do not
have to care about any skewness. The number of points is large enough
to provide a good estimate of the mode with the median. In order to
find the best radius $x$ for the sky level computation, we plotted the
luminosity profile and chose a range where we could not detect any
statistically significant deviation from a constant. Since spiral
galaxies fall exponentially, 1 isophotal radius turned out to be
a fairly good choice for the beginning of such range.

It is obvious that flat fielding--which we obtained from twilight
exposures--is critical and much care must be exercized in making sure
that no large--scale components remain. Fortunately, small--scale
variations are smoothed out by the azimuthal integration. When the sky
level is computed from pixels located at all position angles within a
chosen background region, the same smoothing is {\it de facto} applied.

\subsubsection{Arm Rejection}

One of the aims of this investigation was to isolate the different
populations superimposed on the disks of spirals. This means that we
had to mask the extreme population I component made out of gas and
young stars belonging to the arms. The integrated signal from these
stars is bluer (and brighter) than the inter--arm population and can
be removed from the pixel list via color {\em and} brightness
criteria.  To this purpose a color image was computed from $V'$ and
$I$ images. Here $V'$ \ is a synthesized $V$ band made from the sum of
all four filter bandpasses, which has an effective wavelength close to
that of the standard $V$ system. The flux criterion by which only blue
pixels {\em brighter} than a pre-assigned value was used to prevent
low S/N pixels from being rejected. Such faint pixels could have a blue
color entirely due to noise-induced fluctuations and would tend to
introduce a bias toward the red. Knapen \& Beckman (1996) 
\markcite{KB96}
used an H$\alpha$ image in order to outline star formation regions
(see their Figure 5). We believe our method to be as good as theirs
because it operates at the smaller pixel scale. Nevertheless, the
criticallity of the rejection criteria is somewhat alleviated by the
conclusion that the arms covered a rather small surface compared to
the disk. Beckman et al. (1996) \markcite{BPKCG96} propose a restricted
version of the present method by rejecting pixels having a flux 7\%
higher than the mean $I$ band value at a given radius. This rejection
process, based only on a flux criterion, is probably not as good in
general as the one based on H$\alpha$ image because it can lead to
systematic under--rejection in annuli intercepted by quasi-circular
arms.

\subsection{Calibrating the Indices into the Lick System}

So far we have built a 2-D collection of spectral indices in an
instrumental system which would depend on a number of factors. These
are the filters' design, small velocity-dependent differences between
the bandpasses of the filters and the rest--frame bandpasses of the
galaxy, and atmospheric effects during the observations, such as
transparency differences between exposures taken through the various
filters composing an index. We then proceeded to establish the
transformation to the system defined by the Lick standard stars of
FFBG85\markcite{FFBG85}. This multi--step procedure consists of the
following. (a)~Lick standard stars were observed spectroscopically and
their observed energy distributions were put into the standard flux
system through the use of flux standards observed nightly (\S
4.2.1). Our filter bandpasses, shifted to zero velocity, were then
integrated numerically on the resulting energy distributions.
Notice that the fluxing procedure assured night-to-night stability. In
addition, we used Kennicutt's library of galaxy SEDs (Kennicutt
1992\markcite{K92}) to test the index behavior with empirical
composite populations. (b)~Spectroscopy of the bright central region
of our program galaxies, including two ellipticals, were obtained
during the same run and their fluxed energy distributions were
integrated numerically within the same filter bandpasses as the
standard stars, but shifted to the rest-frame of the galaxy (see \S
4.2.2). (c)~Some additional corrections due to the difference in
spectral resolution between the stellar and galactic observations, as
well as possible velocity effects, were necessary (see \S 4.2.2). We
next describe each step in detail and give a succinct listing of all
steps at the end of \S 4.2.4. .

\subsubsection{Use of Stellar Observations and Published SEDs}

In January, March and June 1995 and in June 1996 we observed
spectroscopically 35 stars in common with FFBG85 and covering a wide
range in spectral type and metallicity\footnote{\footnotesize Their HR
catalogue numbers are as follows: 1346, 1373, 2478, 2600, 2697, 2821,
2854, 3461, 3905, 4365, 4521, 4932, 5227, 5270, 5370, 5480, 5681,
5744, 5826, 5854, 5888, 5901, 5940, 5947, 6014, 6018, 6064, 6103,
6136, 6159, 6299, 6770, 6817, 6872, and 7148.}.

Figures~2 and 3 show the standard (see FFBG85\markcite{FFBG85}) 
versus
observed values (filters' definitions) of the indices, which are well
represented by straight lines. Linear regressions of the stellar data
yield the following relations:
\begin{equation}
{\rm Mg_2}_{\rm Lick,\;std} = 1.18(\pm 0.03){\rm Mg_2}_{\rm OMM} 
+ 0.029(\pm
0.006)\;\;\;{\rm [mag]}
\end{equation}
\begin{equation}
{\rm Fe5270}_{\rm Lick,\;std} = 0.67(\pm 0.05){\rm Fe5270'}_{\rm 
OMM}
+ 1.1(\pm 0.2)\;\;\;{\rm [\AA]}
\end{equation}
where numbers between parenthesis correspond to a $1\sigma$ error on
the respective coefficient. In these equations, it is important to
emphasize that the Mg$_2$ and Fe5270$'$ indices used correspond {\em
both} to the OMM definitions, i.e. they match the OMM filters'
bandpasses. The above slopes were essentially identical when
derived from the composite populations of a  sample of galaxies with
published indices which included some of Kennicutt's spectra and two
ellipticals and a S0/Sa observed by us, and when derived from stars. 

%\placefigure{fig2}

%\placefigure{fig3}

%\placefigure{fig4}

As expected, the slope of the equation for Mg$_2$ is close to unity,
and Figure~2 shows a tight relation, with a dispersion around the mean
line of $\sigma \sim$ 0.03 {\tt mag}, which includes stars and
galaxies. Also as expected, the slope of the equation for Fe5270
differs significantly from unity (Figure~3), and the uncertainty in
the slope is larger. The slopes for stars and galaxies are consistent
with each other. The Fe5270 plot displays a much larger dispersion
than the one for Mg$_2$, of $\sigma \sim$ 0.2 \AA. Because the large
dispersion around the mean lines for the Fe5270 plots might suggest an
effect due to our unorthodox pseudo-index definition, we tested the
intrinsec dispersion of the Lick indices by studying the residuals of
the values obtained from our high S/N stellar sample when using the
standard index definition. Figure~4 shows the histogram of residuals
derived by comparing the published indices to the measured values. We
obtain an intrinsic dispersion comparable to that of Equation~11
suggesting that the Fe5270 standard system itself is not in general as
well determined intrinsically as the Mg$_2$ one. We conclude that the
standard Fe5270 system can be fairly well reproduced from our filter
definition (Fe5270$'$) after correction for the slope indicated above.

%\placefigure{fig5}

The imaging Mg$_2$ index and the imaging Fe5270$'$ pseudo-index were
then multiplied by their derived transformation slopes to put the
observations of our galaxy (M~100) in the Lick system to within an
additive constant which is derived below.

\subsubsection{Zero Points}

All spectroscopic observations for galaxies were obtained with a slit
opened to 6\arcsec .1 in order to increase the S/N, and to render
accessible a large image section to the zero point adjustement between
the spectroscopy and the imagery. This slit width corresponds to a
resolution of 34 \AA \ FWHM, far lower than the value used in most
studies, and should produce a significant offset with respect to the
Lick system.

In order to determine this index offset high-resolution stellar
spectra were degraded to match the galaxy spectral resolution and the
resulting differences were added to the galaxy indices. In addition
the fluxed--Lick correction of 0.017 {\tt mag}, adopted by Gonz\'alez
et al. (1993) \markcite{GFW93} was added to the Mg$_2$. This combined
corrrection was quite consistent with the intercept of the Lick versus
M\'egantic plot for stars, once the M\'egantic stellar observations
were placed themselves into the standard Lick IDS dispersion of 8 \AA
\ FWHM.
\begin{equation}
{\rm Mg_2}_{\rm Lick,\;computed} = {\rm Mg_2}_{\rm OMM} + 0.018 
+
0.017\;\;\;{\rm [mag]}
\end{equation}
\begin{equation}
{\rm Fe5270}_{\rm Lick,\;computed} = {\rm Fe5270}_{\rm OMM} +
1.307\;\;\;{\rm [\AA ]}
\end{equation}
where the indices Mg$_2$ and Fe5270 follow the Lick definitions.
  
The zero points of the image system were derived by matching the
spectroscopic standard indices for the bright central part of the
program galaxies to those derived from integration of our filter
bandpasses within the exact same region.

\subsubsection{Velocity Broadening and Rotational Corrections}

Spectral broadening due to velocity dispersion differences across a galaxy 
can introduce systematic effects in the determination of spectral indices, in 
particular for Fe5270. Since we are using relatively wide bandpasses in this 
investigation this effect should not be significant; we have used a spectrum to simulate the effect of a bulge-disk extreme difference of 150~${\rm 
km}\:{\rm s}^{-1}$ and found a difference of less than 1\% which we 
disregarded.

 As already discussed we have chosen nearly face-on spirals (such as
M~100) for our extended program for a number of reasons. Arm-interarm
separation is best achieved on these objects because we have more
available pixels for a given distance, reddening corrections (see
below) are less significant, and differential Doppler effects are
clearly smaller. Very large spirals with rotational speeds of order
300~${\rm km}\:{\rm s}^{-1}$ will shift the effective rest--frame
bandpasses of our filters by as much as 5 \AA\ if seen edge-on. In
practice this effect, which can be accounted for by using a suitable
energy distribution to compute a correction, will be small and in most
cases with $i \le {\rm 45} \deg $ the $\lambda$ shift will be less
than 6\% of the bandpass width. For NGC~4321 it is only 1.9 ${\rm \AA
}$ and since a spectral simulation shows that the impact on the indices is 
negligible (well below 1\%) we have not computed a correction. Notice 
also that
uncorrected azimuthal integrations will always result in a ${\em
broadening}$ of the bandpass by less than twice the $\lambda $ shift
when averaged over an annulus.

\subsubsection{Reddening Corrections}

In order to examine the effect of internal absorption on our indices,
we have used a representative energy distribution, the nuclear
one, and applied different amounts of visual extinction. The results
are shown in panels (a) and (b) of Figure~5. Notice the non-monotonic
behavior of Mg$_2$, which reaches a maximum correction of $\sim$ 4\% 
near
$A_{V}$ = 5 mag. Since we do not expect values of $A_{V}$ larger than
2--3 mag, we did not apply a reddening correction to our data. We do
not expect any significant contribution to an index gradient in the
disk of M~100 from a gradient in the dust distribution.

%\placefigure{fig6}

Let us summarize the contents of \S 4.  We: (1) Obtained the
instrumental (OMM)--to--standard (Lick) transformations (Equations~10
and 11 of \S 4.2.1). (2) Measured the indices on the spectra of the
bright central regions of the target galaxies using the standard
(Lick) filter definitions (\S 4.2.2). (3) Computed and added the
dispersion corrections for both indices, and the flux correction for
Mg$_2$; this placed the values of the indices of the bright central
regions in the Lick system (Equations~12 and 13 of \S 4.2.2). (4)
Measured the indices on the images, which were in the OMM system (\S
4.2.1). (5) Applied the slopes found in (1) in order to transform the
OMM filter system into the Lick system. (6) Computed the offsets
between the spectroscopic values (from 3) and the imaging values (from
5) (\S 4.2.2). (7) Obtained the Lick indices at any point along the
galaxy profile by multiplying their values on the OMM system by the
slopes found in \S 4.2.1 and adding the offsets found in \S 4.2.2.

\section{First Results for M~100}

We start by showing a sequence of images of M~100, none of which has 
been deprojected, and all of which have the same orientation and scale. 
Figure~6 shows this
galaxy in $V'$ light, an image obtained by coadding all four narrow
band filters. Figure~7 provides an image formed by the $V'-I$ color;
this was the reference image for the arm-interarm discrimination
described in \S 4.1.3. Notice in Figure~7 the presence of the well
known ring at a radial distance of 10\arcsec \ to 22\arcsec \
(Arsenault et al. 1988\markcite{ABGR88}), and the bar extending to
about 60\arcsec (Knappen et al. 1996\markcite{}).  Figure 8 shows M~100 
again in $V'$ light, as in
Figure~6, but with the arm pixels and unwanted objects masked.
Figures~9 and 10 show the {\em unmasked} image of M~100 in the 
Mg$_2$
and in the Fe5270$'$ index, respectively. In these last two images, which 
have
been smoothed with a $17\times 17$ pixel square box, stronger indices
are coded brighter. Notice that the bar is apparent in Figure~7, but
not in figure~9 or 10, suggesting that the bar indices matche those of the 
disk
within the bar radius. There is clear evidence of a Mg$_2$ gradient,
shown in panel (a) of Figure~11 which reaches to about 4
exponential scale lengths. Notice that we do not attempt here or elsewhere 
in this paper to decompose the bulge/disk contributions. However, 
examination of the profile decomposition by Kodaira et al. 
(1986)\markcite{KWO86} shows that the bulge dominates the observed 
surface brightness distribution only within the central 1--2 kpc (and again 
past 24 kpc).  Panel (b) of the same
figure shows little evidence, within the errors, of a Fe5270 gradient
except perhaps within the inner 5 kpc (the scale is assumed to be 1 kpc
= 12\arcsec, see Table~4). We repeat the above figure in Figure~12, where the horizontal axis is now in units of $V$ surface brightness (the latter
from Beckman et al. 1996\markcite{BPKCG96}). Near the center the 
profiles are probably dominated by the central ring (Arsenault et al. 
1988\markcite{ABGR88}).

%\placefigure{fig7}

%\placefigure{fig8}

%\placefigure{fig9}

%\placefigure{fig10}

%\placefigure{fig11}

%\placefigure{fig12}

There is some indication in Figures 11 and 12 of a change in the Mg$_2$
gradient near R=5 kpc, which could be due to the effect of the well known
bar present in M~100 (see Figure 7). Bars are believed to be
efficient carriers of disk material of lower mean abundance from
outside the co--rotation radius into the central regions (Friedli \&
Benz 1995\markcite{FB95}; Courteau et al. 1996;  Beauchamp, Hardy \& 
Friedli 1997, in preparation). 

Although we refrain from assigning separate Mg and Fe abundances to
M~100 within the context of this paper, we use the results of Idiart
et al. (1996) \markcite{IFPDC96} (their Equation 4) to transform
Figures~11 and 12 into [Mg/Fe] gradients under certain assumptions. Their 
models are valid for bulge populations, but show a similar evolutionary 
time  dependence for the {\rm Mg/Fe} and Fe5270 indices, dependence that 
drops out of the model when the [Mg/Fe] {\em ratio} is computed. Thus, 
their computation of this
fundamental ratio, for bulges, is {\em independent of the mean age of the
population}, and is represented by
\begin{equation}
[{\rm Mg/Fe}]= -0.042 -0.2{\rm Mg}_2 - 0.063 <{\rm Fe}> - 7.73 ({\rm
Mg}_2/<{\rm Fe}>)\;\;\;{\rm [dex]}
\end{equation}
which assumes a minimum time of 1 Gyr for the onset of type Ia
supernovae. Here $<$Fe$>$ is the mean strength (in \AA ) of the Fe5270
and the Fe5335 features. Since we have not measured Fe5335, we have
collected together the observations of spiral bulges from Idiart et
al.  (1996) \markcite{IFPDC96}, and Jablonka et al. (1996)
\markcite{JMA96} to establish a relationship between $<$Fe$>$ and
Fe5270; the resulting linear correlation, derived from 79 bulges was
\begin{equation}
<{\rm Fe}> =0.91\; {\rm Fe5270} + 0.091\;\;\;{\rm [\AA ]}
\end{equation}
Introducing this relation into the previous one provides the results
shown in Figure~13. There the error bars represent 1$\sigma $ of the
combined errors of the Idiart et al. 1996\markcite{IFPDA96} relation
and those of the $<$Fe$>$ versus Fe5270 calibration. If one assumes a 
radial-independent SNe onset time, then the above computations may 
represent a reasonable approximation to the disk [Mg/Fe] ratio. This is, 
however, probably unrealistic, and it is likely that Figure~13 is valid only 
for the central regions of M~100, where we see a systematic overabundance  of $\alpha $-elements.  Observations of a large sample of spirals,  as well as 
more realistic calibrations of the individual Mg and Fe
abundances, will be
combined in future papers to obtain a more conclusive picture of the
distribution of stellar abundances and abundance ratios across
spirals.

%\placefigure{fig13}

\acknowledgments

D. B. was partly financed by an NSERC--Canada graduate fellowship.
E. H. acknowledges support from an NSERC--Canada research grant. This investigation was also supported by grants from
FCAR--Qu\'ebec. The authors wishes to express their gratitude to the
OMM staff, in particular B. Malenfant and G. Turcotte, for their help
during the observations.

\appendix
\section{Error Computations For Mg$_2$ And Fe5270}

We assume, for a given function $u=f(x,y,...)$, where $x$ and $y$ are
independant variables, that:
\begin{equation}
\sigma^2_u=\left(\frac{\partial f(x,y,...)}{\partial
x}\right)^2\sigma^2_x+ \left(\frac{\partial f(x,y,...)}{\partial
y}\right)^2\sigma^2_y + ...\simeq (\Delta u)^2
\end{equation}
 Thus, for Mg$_2$, the function is
\begin{equation}
{\rm Mg_2}=-2.5\log \left( \frac{{\cal F}_{\rm Mg_2}} {{\cal F}_{\rm
C}} \right)
\end{equation}
and we have that
\begin{equation}
\sigma^2_{\rm Mg_2}=1.086^2\left(
\frac{\sigma^2_{\cal F_{\rm Mg_2}}}
{{\cal F}_{\rm Mg_2}^2}+
\frac{\sigma^2_{\cal S_{\rm Mg_2}}}
{{\cal F}_{\rm Mg_2}^2}+
\frac{\sigma^2_{\cal F_{\rm C}}}
{{\cal F}_{\rm C}^2}+
\frac{\sigma^2_{\cal S_{\rm C}}}
{{\cal F}_{\rm C}^2}
\right)
\end{equation}
Where $\cal S$ denotes the sky level. The standard error of the measurement are given by $\sigma_{\rm
mean}=\sigma / \sqrt{n-1}$:
\begin{equation}
{\sigma_{\rm Mg_2}}_{\rm mean}=1.086\sqrt{\frac{\left(
\frac{\sigma^2_{\cal F_{\rm Mg_2}}}
{{\cal F}_{\rm Mg_2}^2}+
\frac{\sigma^2_{\cal S_{\rm Mg_2}}}
{{\cal F}_{\rm Mg_2}^2}+
\frac{\sigma^2_{\cal F_{\rm C}}}
{{\cal F}_{\rm C}^2}+
\frac{\sigma^2_{\cal S_{\rm C}}}
{{\cal F}_{\rm C}^2}
\right)}{n-1}}
\end{equation}

The derivation and the assumptions are the same for Fe5270.
The corresponding function is
\begin{equation}
{\rm Fe5270}=55 \left[ 1- \left( \frac{{\cal F}_{\rm Fe5270}}{{\cal
F}_ {\rm C_{\rm w}}} \right) \right]
\end{equation}
We find that
\begin{equation}
\sigma^2_{\rm Fe5270}=\left(\frac{55}{{\cal F}_{\rm C_w}}
\right)^2\left(\sigma^2_{{\cal F}_{\rm Fe5270}} +
\sigma^2_{{\cal S}_{\rm Fe5270}}+\frac{\sigma^2_{{\cal F}_{\rm 
C_w}} + 
\sigma^2_{{\cal S}_{\rm C_w}}}{{\cal F}_{\rm C_w}}
\right) 
\end{equation}
which gives
\begin{equation}
{\sigma_{\rm Fe5270}}_{\rm mean}=\frac{\left(\frac{55}{{\cal F}_{\rm
C_w}}\right)\sqrt{ \sigma^2_{{\cal F}_{\rm
Fe5270}}+ \frac{\sigma^2_{{\cal F}_{\rm C_w}}+ \sigma^2_{{\cal 
S}_{\rm
C_w}}} {{\cal F}_{\rm C_w}^2}}}{\sqrt{n-1}}
\end{equation}

%%%%%%%%%%%%%
%% REFERENCES
%%%%%%%%%%%%%

\newpage

\figcaption[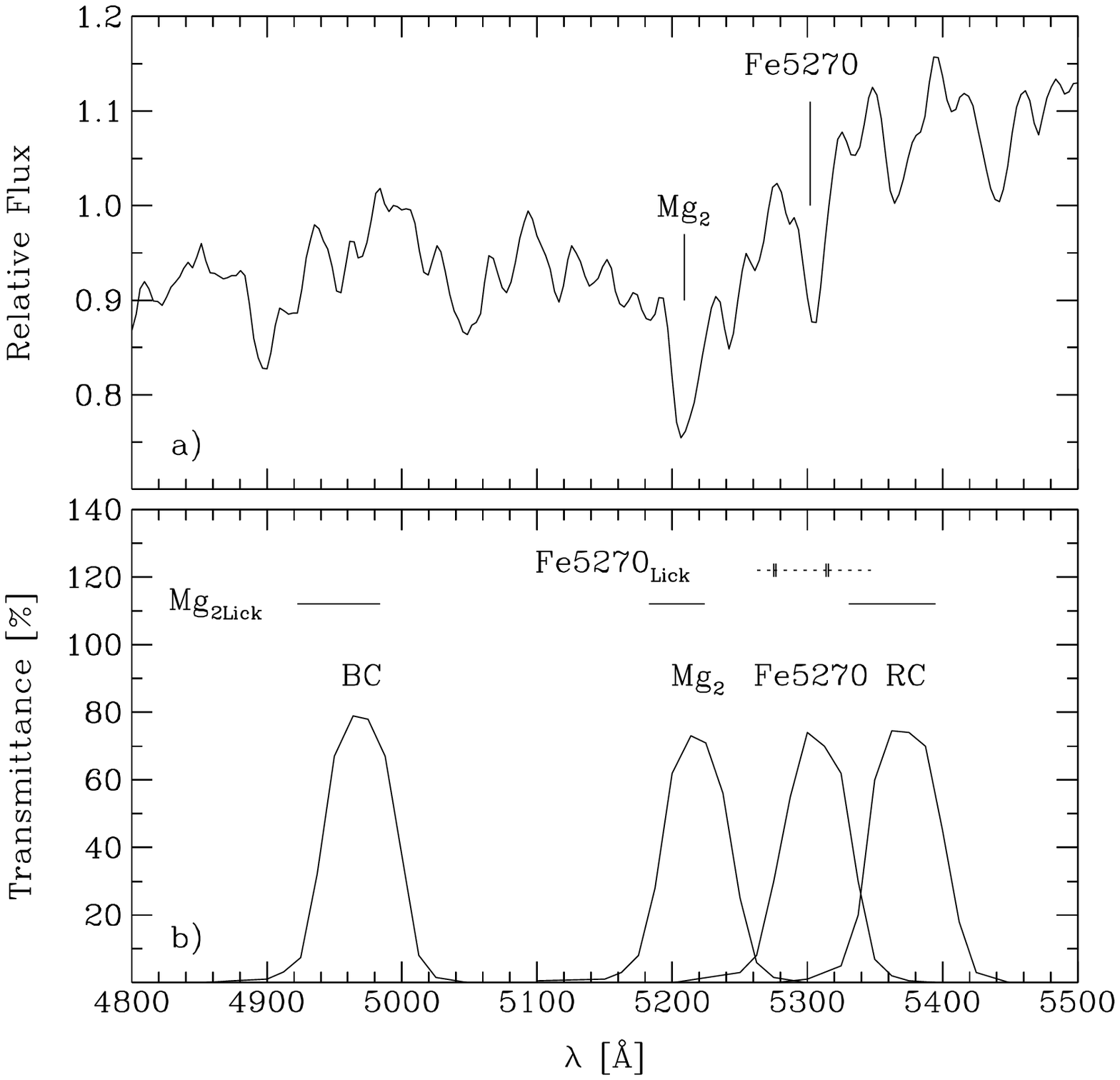]{Top panel: K star SED redshifted to 2000 
${\rm km}\:{\rm s}^{-1}$ to match the filter set shown below. The Mg 
and Fe features measured in this investigation are identified. Bottom panel: 
The transmission curves of our filter set at 23~$^{\circ}$C. The BC and 
RC labels identify the blue and red continua respectively. The solid 
horizontal lines represent the three Lick bandpasses that define the standard 
Mg$_2$ index; the discontinuos horizontal lines those of the standard 
Fe5270 index.\label{fig1}}

\figcaption[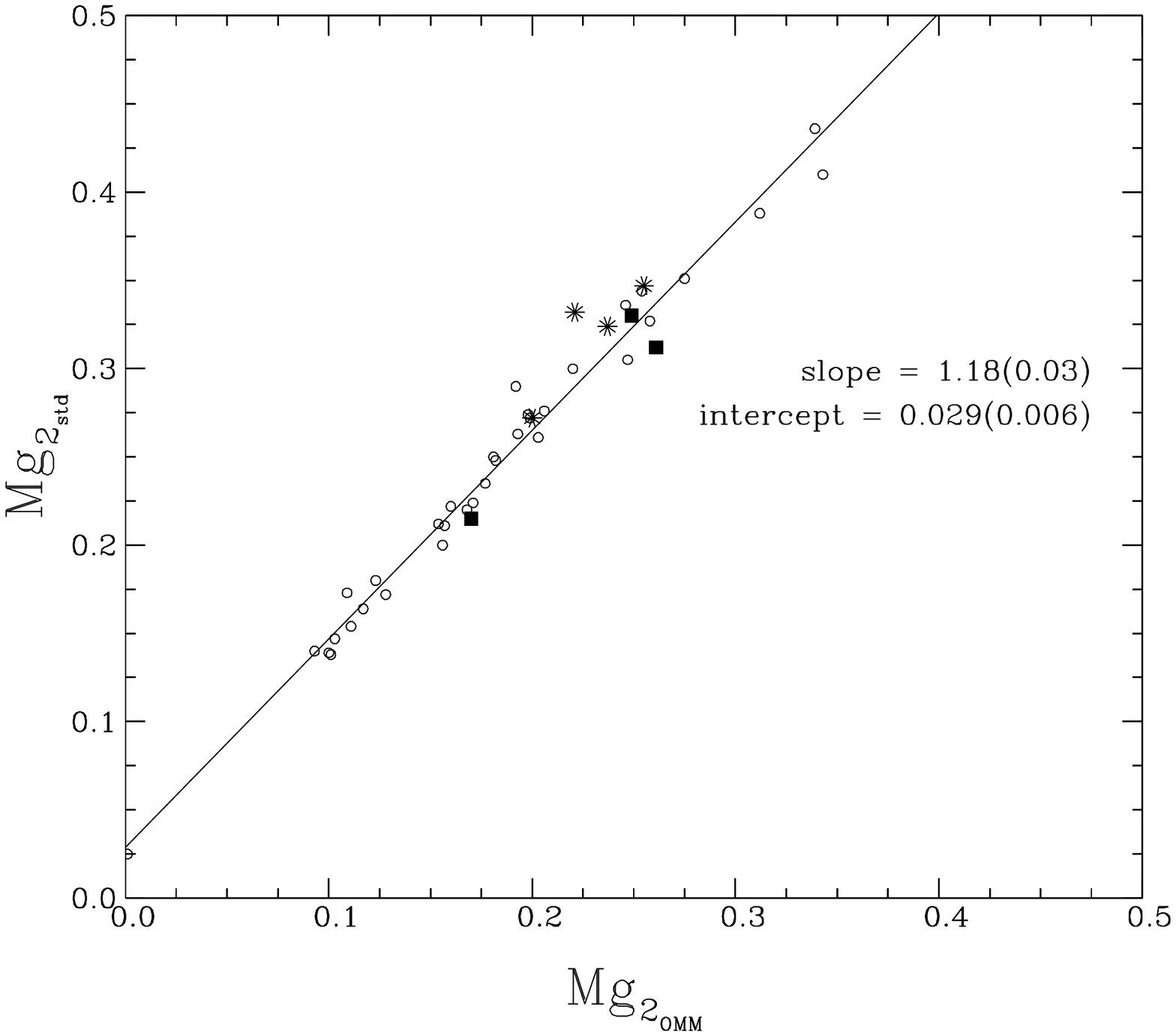]{Transformation between standard (Lick)
and measured Mg$_2$ index (OMM filter definition) derived from 
integration of the stellar spectra (open circles), of Kennicutt's galaxies
(asteriscs), and of the three galaxies of our program (NGC 2655, 2974
and 4473; solid boxed) which have published standard Mg$_2$ indices . 
The least square parameters of the fit are indicated, with their errors 
within parenthesis. \label{fig2}}

\figcaption[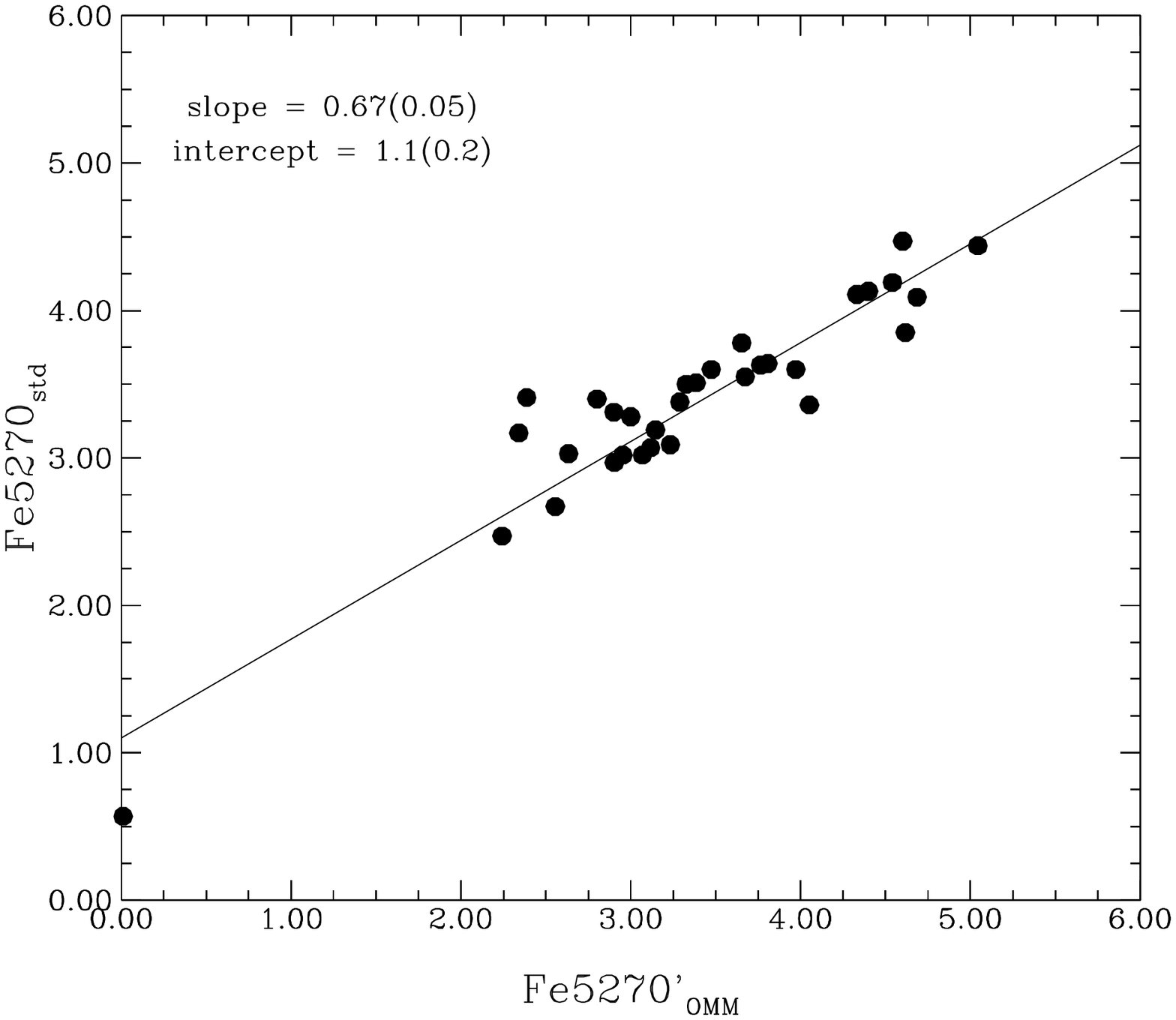]{Same as Figure~2, but for the stellar 
Fe5270 data. \label{fig3}}

\figcaption[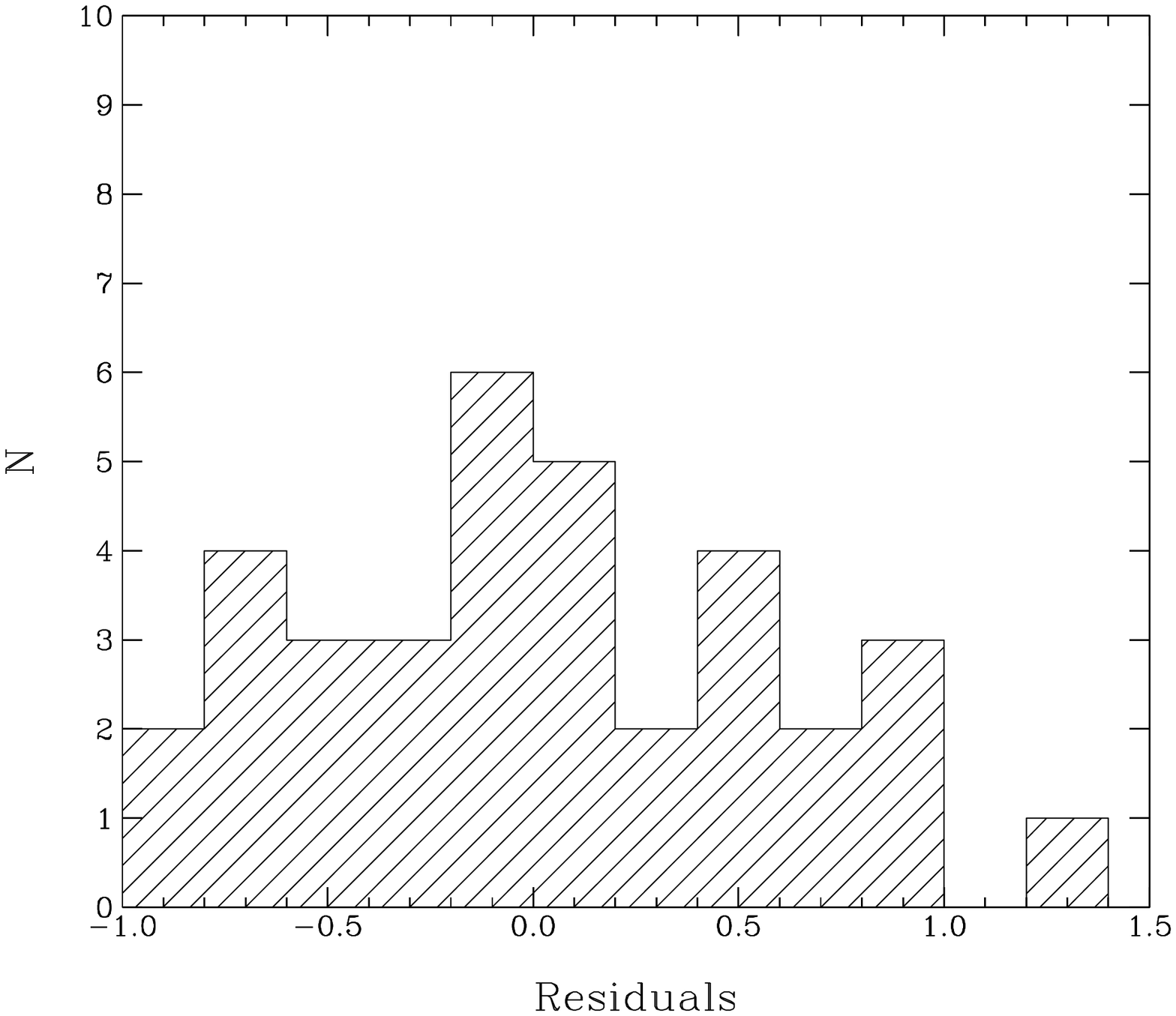]{Histogram of differences between 
published Lick Fe5270 values (from FFBG85) and measured Fe5270 
values.\label{fig4}}

\figcaption[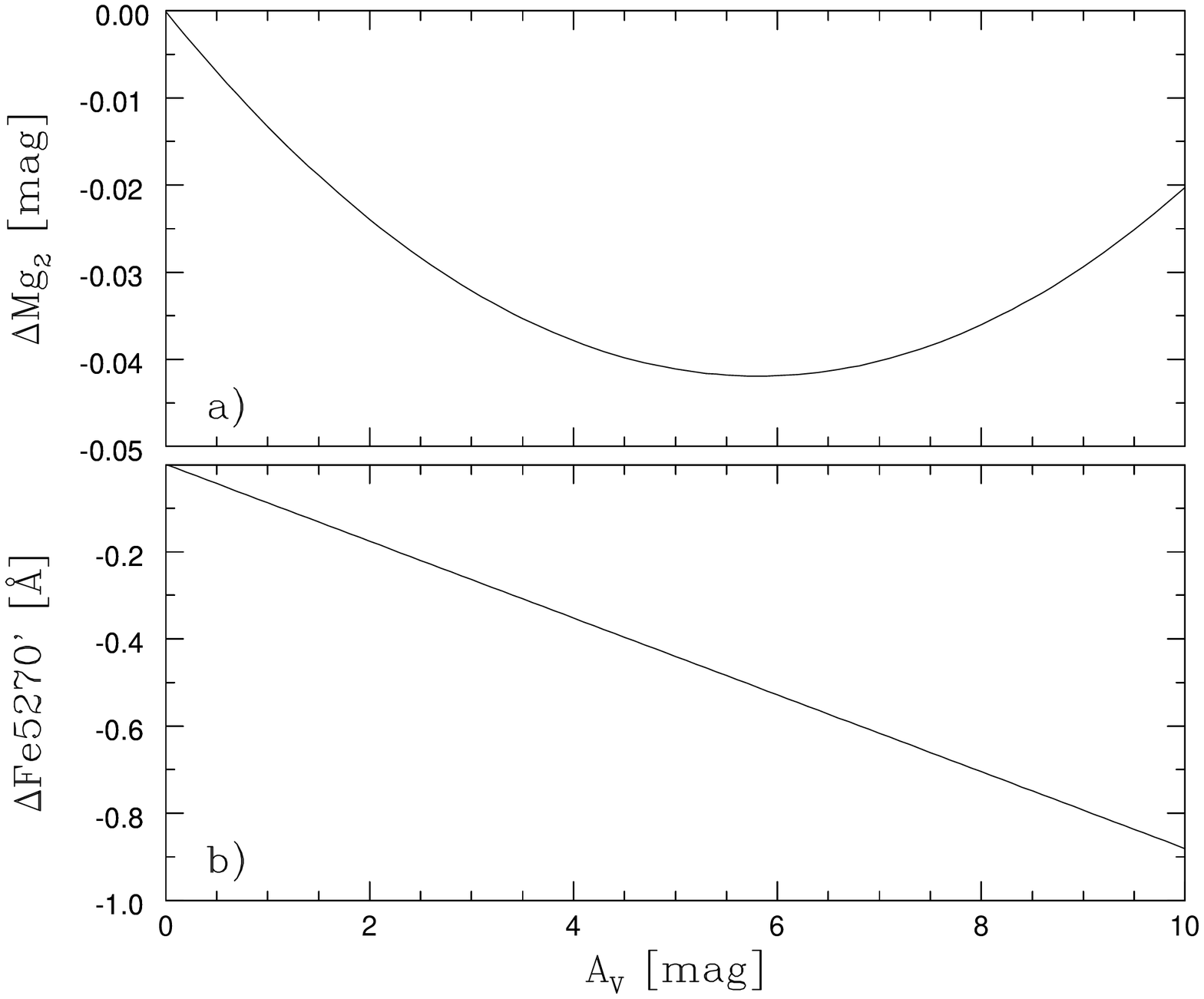]{The dependence of index values on internal 
absorption as simulated using the measured central spectrum of 
M~100.\label{fig5}}

\figcaption[Beauchamp.fig6.ps]{Direct image of the central 13\arcmin\ of 
M~100 in $V'$ band. The reference horizontal line is 1\arcmin\ long, and 
the circle has a radius corresponding to 1 isophotal radius (R25 = 
222\arcsec). North is up, east to the left. All five images of M100 (see 
below) have the same scale and field of view, and are 
unprojected\label{fig6}}

\figcaption[Beauchamp.fig7.ps]{M100 in $V'-I$ color.\label{fig7}}

\figcaption[Beauchamp.fig8.ps]{Same as Figure~6 but with arms and
bright stars masked.\label{fig8}}

\figcaption[Beauchamp.fig9.ps]{The Mg$_2$ index unmasked ``image'' of 
M~100.
The original image has been median filtered with a $17\times 17$ pixel
square box.\label{fig9}}

\figcaption[Beauchamp.fig10.ps]{Same as Figure~9 but for the Fe5270$'$
index.\label{fig10}}

\figcaption[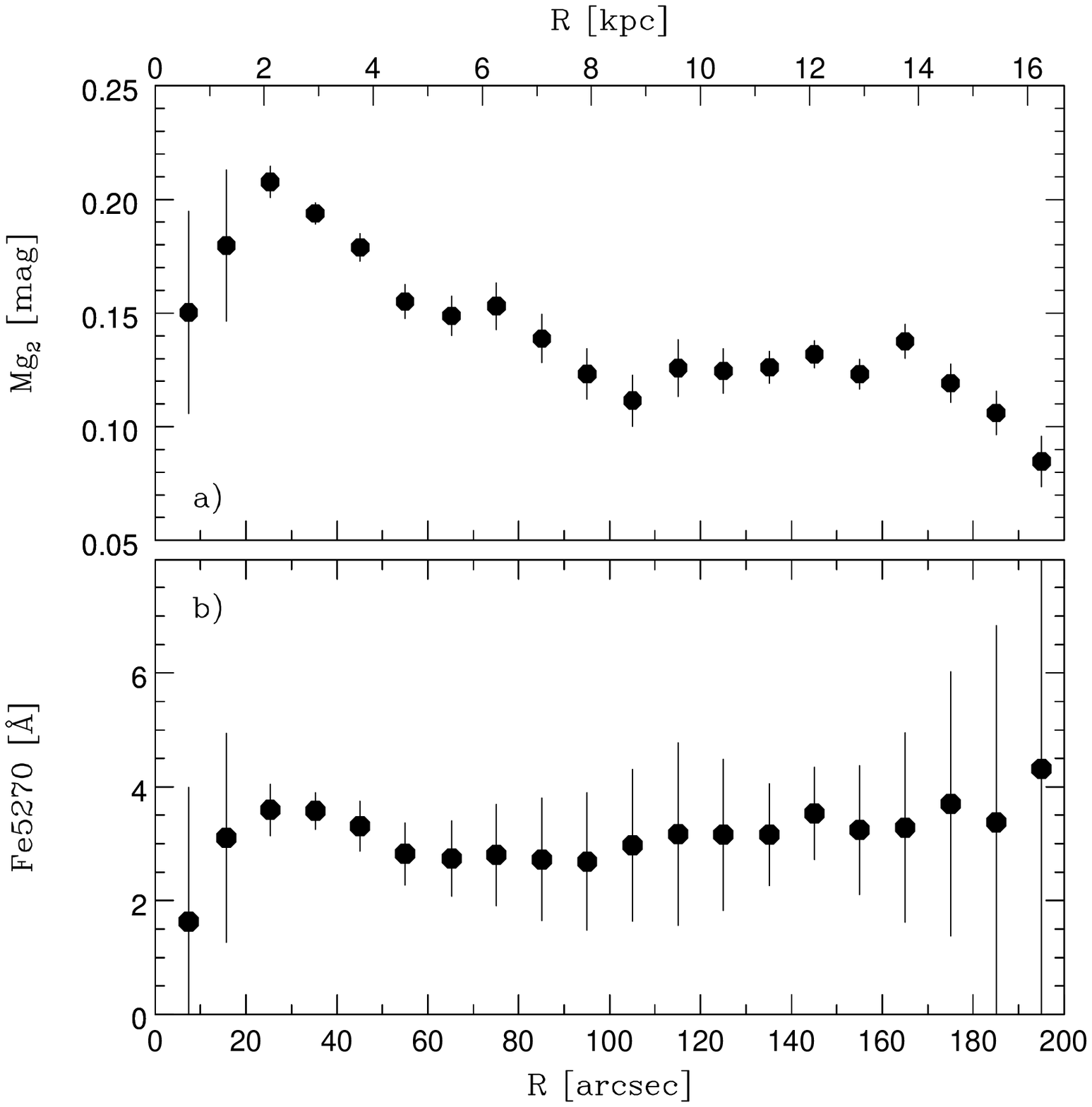]{Index profiles as function of deprojected 
radial distance in arcsecs and in kpc.\label{fig11}}

\figcaption[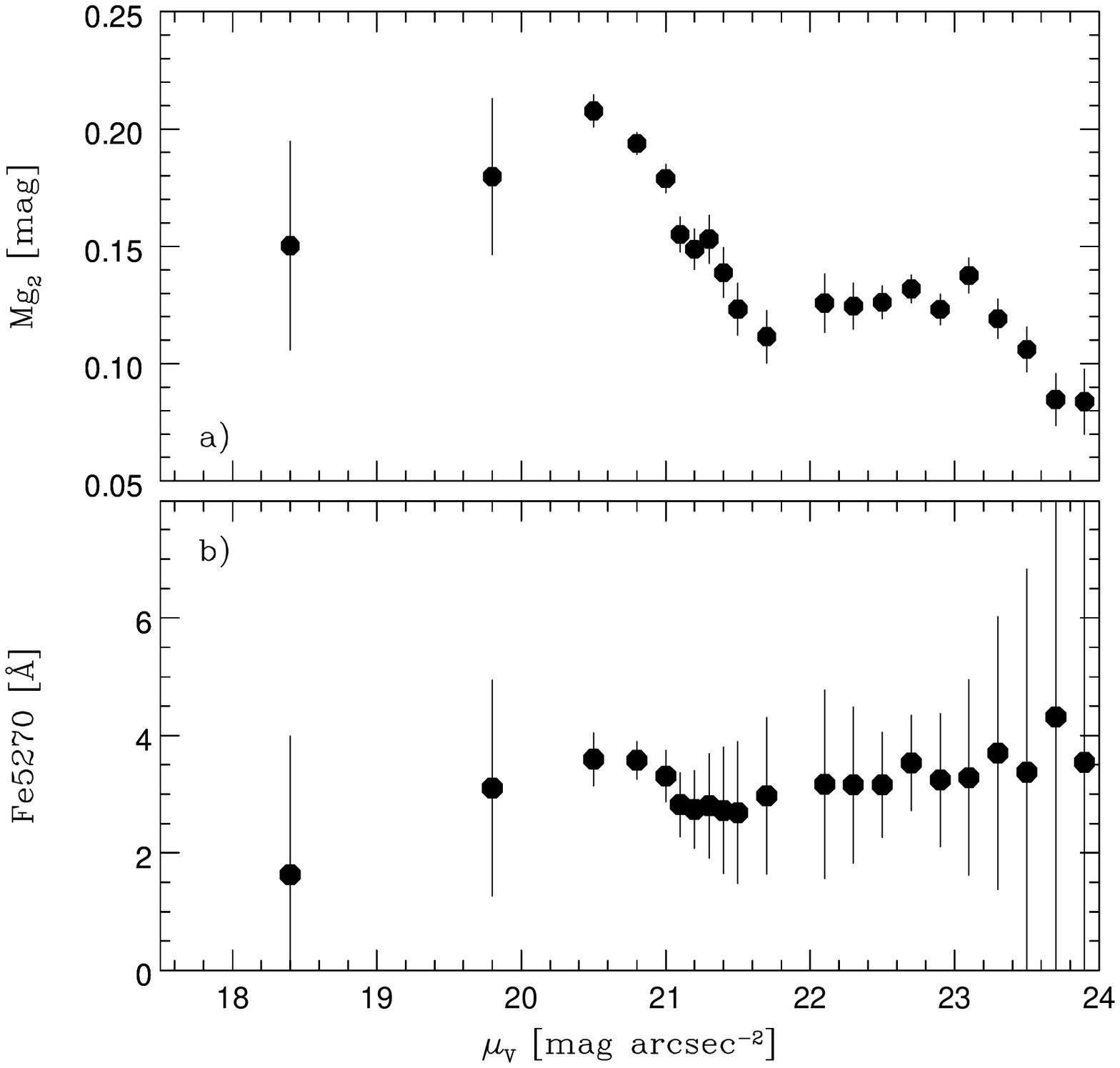]{Index profiles as function of V surface
brightness.\label{fig12}}

\figcaption[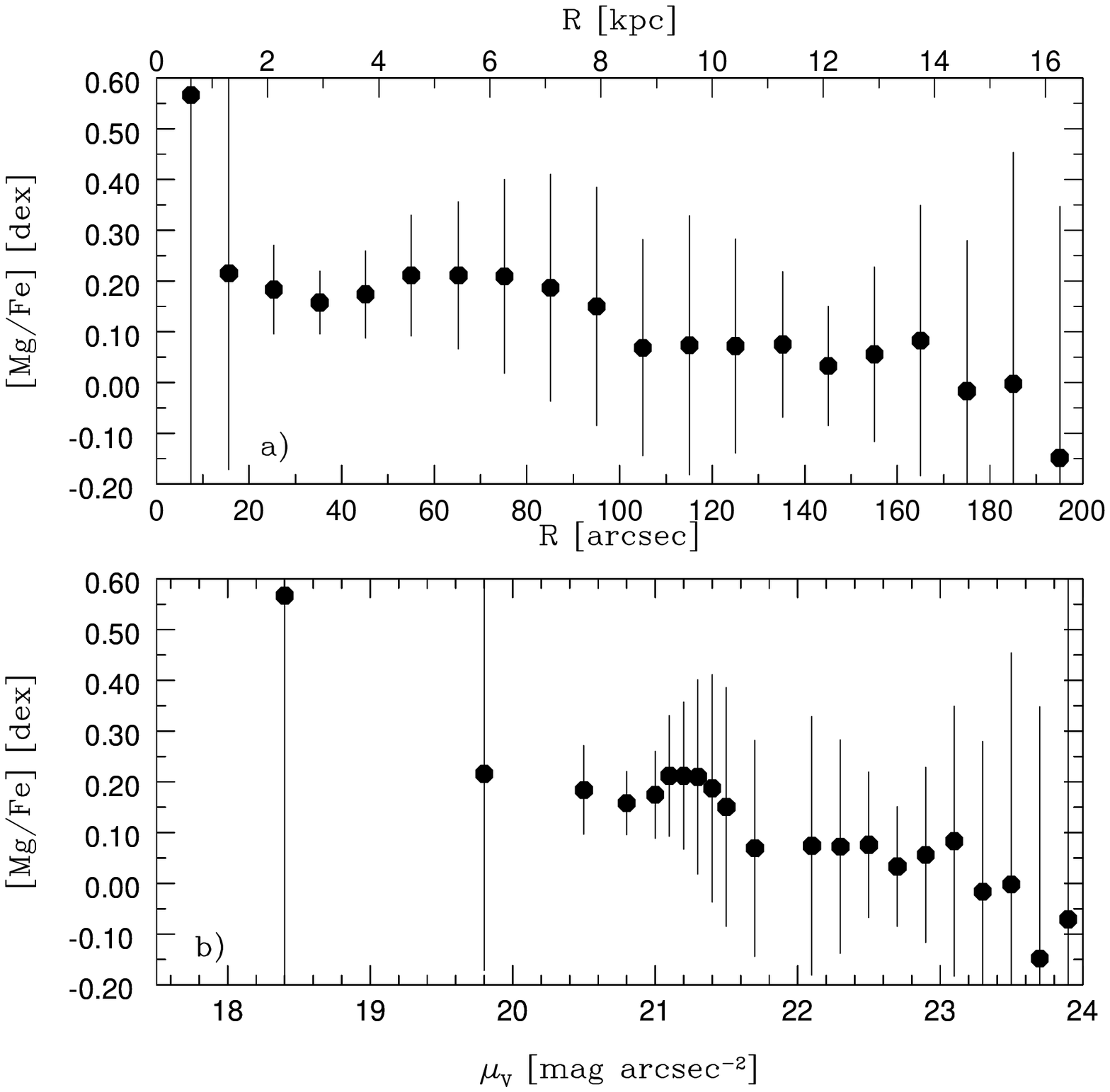]{Model abundance ratio [Mg/Fe] as 
function of radial distance and V surface brightness.\label{fig13}}

\end{document}